# Atom phase controlled noise-free photon echoes

Byoung S. Ham
School of Electrical Engineering and Computer Science, Gwangju Institute of Science and Technology
123 Chumdangwagi-ro, Buk-gu, Gwangju 61005, S. Korea
bham@gist.ac.kr

Abstract: Rephasing in photon echoes is a fundamental mechanism of retrieving optical information stored in a collective ensemble of atoms or ions. With an extremely weak quantum optical data, population inversion by the rephasing process is inevitable resulting in serious quantum noise. Here, an inversion-free rephasing method in photon echoes is presented by using deshelving-based atom phase control. The present method makes photon echoes directly applicable to quantum memories with benefits of extended photon storage time and nearly perfect retrieval efficiency. In this revised version only the rephasing pulse propagation direction is corrected to make a silent echo of E1 in Fig. 1.
(Version is uploaded on Jan. 28, 2011; Version 2 is uploaded on Nov. 14, 2016)

Optical rephasing of photon echoes is the preeminent mechanism of photon storage in a collective ensemble of atoms[1]. Even with the benefits of a multimode storage capability in the time domain and a nearly perfect retrieval efficiency in a backward propagation scheme, however, inevitable population inversion by a π optical rephasing pulse results in serious quantum noise[2,3]. To remove population inversion-caused spontaneous emission noise by the optical rephasing pulse, several modified photon echo protocols such as atomic frequency comb[4-7] and gradient echoes[8-10] have been presented recently for quantum memory applications. Here, an atom phase control technique using double rephasing and optical deshelving is presented for spontaneous emission-free photon echoes. As a result, a spontaneous emission noise-free storage time-extended quantum memory protocol is obtained. Unlike photon echoes or other modified echoes reversely ordered and exponentially decayed[1-10], the present method gives an additional benefit of exponential decay-free normally ordered photon echoes.

For quantum network applications[11], a quantum interface between photons and a matter has been studied for remote quantum communications[12,13]. Quantum optical memories in solids for these applications have drawn much attention recently[6,7]. From distributed quantum computing to long-distance quantum communications, ultralong quantum memory has been an essential element, because ultralong photon storage capability determines implementations of long-distance quantum communications based on quantum repeaters performing cascaded quantum purifications[14,15]. In general the required photon storage time of quantum memories should be longer than a second to satisfy long-distance quantum communications[16]. To extend photon storage time in photon echoes, a technique of control deshelving pulses has been proposed[17].

A quantum memory is a quantum device to store and retrieve an unknown quantum state on demand. Because photons are fragile to environmental interactions, an interface between photons and a rigid solid medium is highly desirable[4-10]. Compared with atomic media such as vapors and cold atoms[12,13], solid media such as rare-earth doped crystals are much more robust in atomic diffusion and decay processes, and advantageous in spectral selections and nonlinearities[18,19]. A liquid helium temperature operation, however, has been a major obstacle in the practical usage of rare-earth doped solids. A recently available electrically pumped closed-cycle cryostat for temperatures near 4 K should remove the low temperature limitations, though. Since the first observations of ultraslow light-based photon storage in atomic media[20], rare-earth doped solids have been intensively studied for potential applications of quantum memories over the last decade in the area of slow light[21], atomic frequency comb echoes (AFC) (refs. 4-7), gradient echoes[8-10], and resonant Raman echoes[23]. The key storage mechanism in these methods is rephasing of inhomogeneously broadened ions, where collective ions play an important role. Compared with a single atom-based quantum memory such as in a single ion storage medium, an ensemble based system provides enhanced photon-atom interactions, achieving a nearly perfect coherence transfer via a full absorption process can be achieved.

Photon echoes have unique intrinsic properties of multimode storage capability in the time domain, contributing to ultrafast information processing[22]. The key storage mechanism of photon echoes is the use of inhomogeneously broadened atoms, where each individual atom phase is controlled to be reversed in the time domain: Rephasing. The rephasing of a photon-excited atom ensemble in a frequency domain enables multiple optical data storage in the time domain, where a maximum writing speed is determined by the inverse of the inhomogeneous broadening of the medium. The photon echo storage time is limited by the given optical phase decay time of the medium, which is much shorter than a millisecond in most rare-earth doped solids[18]. Due to the rephasing of collective atoms or ions, use of photon excited spectral gratings is a fundamental property of photon echoes, benefiting multimode quantum storage in the time domain[4-10]. Here, storage efficiency of optical light inside the medium is obviously not uniform but exponentially decays along the propagation direction according to Beer's law. Thus, the retrieved photons along the same propagation direction must suffer from reabsorption by noninteracted atoms as they propagate to exit the medium[24]. Due to this inherent echo reabsorption problem, photon echo efficiency has been reported normally at less than 1%. To overcome this fundamental limitation in photon echo efficiency, a backward propagation scheme using optical-spin coherence transfer via optical deshelving has been proposed[17], and experimentally demonstrated recently with fifteenfold enhanced retrieval efficiency[25].

The most serious drawback of photon echoes for quantum memory applications is the optical rephasing process-induced population inversion. For classical photon storage, the population inversion could be a benefit of coherent amplification. With quantum information, however, a no cloning theorem prohibits self-amplification of stored light in the optical gain process. In photon echoes especially dealing with an extremely weak light, a nearly complete population inversion by an optical π rephasing pulse results in quantum noise based on both spontaneous and echo-triggered stimulated emissions. Using a suitable propagation scheme as shown in Fig. 1c with a sub-ns optical pulse in a rare-earth doped solid, the spontaneous emission-caused quantum noise can be practically removed. However, the stimulated emission-caused quantum noise cannot be avoided. Thus, direct use of photon echoes has been limited to quantum memory applications[2,3]. To avoid rephasing-caused population inversion, AFC or gradient echo techniques give great benefit by not using the rephasing optical π pulse. Regarding the storage process, however, AFC and gradient echoes should pay for a long preparation time[4-7], lower retrieval efficiency[4-7], or use of dc electric fields[8-10].

In this Letter we present a method to avoid the rephasing-induced population inversion in photon echoes by using all-optical atom phase control with additional control pulses. With all-optical atom phase control, the present technique provides three major breakthroughs to quantum memory applications using photon echoes without paying the extra price: First, spontaneous emission-free near perfect retrieval efficiency; second, normally ordered (rather than reversed) intensity flattened echoes; and third, extended photon storage time. Thus, the present method opens a door to photon echo-based quantum memories with a nearly perfect storage time-extended noise-free multimode photon storage capability.

Figure 1 shows an energy level diagram of the present atom phase controlled photon echoes. In Fig. 1a, a resonant optical field A performs conventional photon echoes with an additional rephasing function, while auxiliary optical field B controls atom phase via optical deshelving. The optical deshelving onto spin state |2> also results in storage time extension due to robust spin decay processes[5,25]. Figure 1b shows a light pulse sequence of Fig. 1a, where E1 represents conventional time-reversed photon echoes, while E2 represents nonreversed storage-time extended photon echoes. Although the primary function of control pulses C1 and C2 is to make a π phase shift to the doubly rephased atoms, resultant coherence transfer to spin states brings storage time extension limited by spin dephasing. This storage time extension can be up to a few tens of milliseconds if an external dc magnetic field is applied[26].

Figure 1c represents a pulse propagation scheme, where the backward propagation scheme implies a near perfect retrieval efficiency as proposed[17] and experimentally demonstrated[25]. The angles among all light beams are chosen to be overlapped inside the medium by more than 90%. Upon dissatisfying phase matching conditions for the first echo E1 with backward rephasing pulses R1 and R2, a silent echo can be formed as shown in Eq. (1)[27]. Because individual atom phase evolutions do not matter with E1, the echo E2' is generated in the same direction as the data D. This echo E2' is now spontaneous emission-free due to no population inversion, and non-reversed in echo sequence (to be discussed in Figs. 2 and 3):

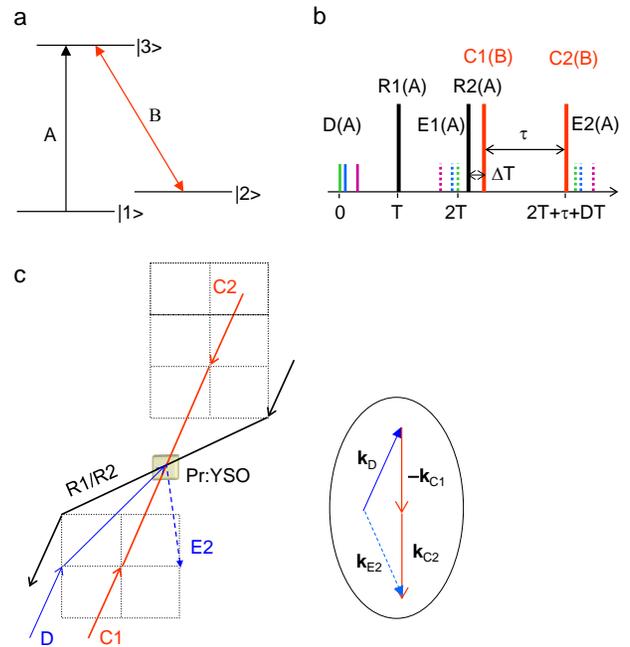

Figure 1. Schematics of atom phase control in photon echoes. a, Energy level diagram. b, Pulse sequence. c, Propagation directions. D: Data pulse; R1/R2: Rephasing pulses; C1/C2: Control deshelving pulses; E2: Echoes.

$$\vec{k}_{E1} = 2\vec{k}_{R1} - \vec{k}_D \approx -3\vec{k}_D. \quad (1)$$
$$\vec{k}_{E2'} = 2\vec{k}_{R2} - \vec{k}_{E1} = \vec{k}_D \quad (2)$$

where $\vec{k}_i$ and $\varpi_i$ are wave vector and angular frequency of the pulse $i$, respectively[25]. For a backward echo E2,

counterpropagating control pulses C1 and C2 are applied right after R2:

$$\vec{k}_{E2} = \vec{k}_D - \vec{k}_{C1} + \vec{k}_{C2}, \quad (3)$$
$$\varpi_{E2} = \varpi_D - \varpi_{C1} + \varpi_{C2}, \quad (4)$$

where $\vec{k}_D = \vec{k}_{E2'}$. Without control pulses, echo E2' cannot be radiated due to its absorptive coherence under the double rephasing scheme.

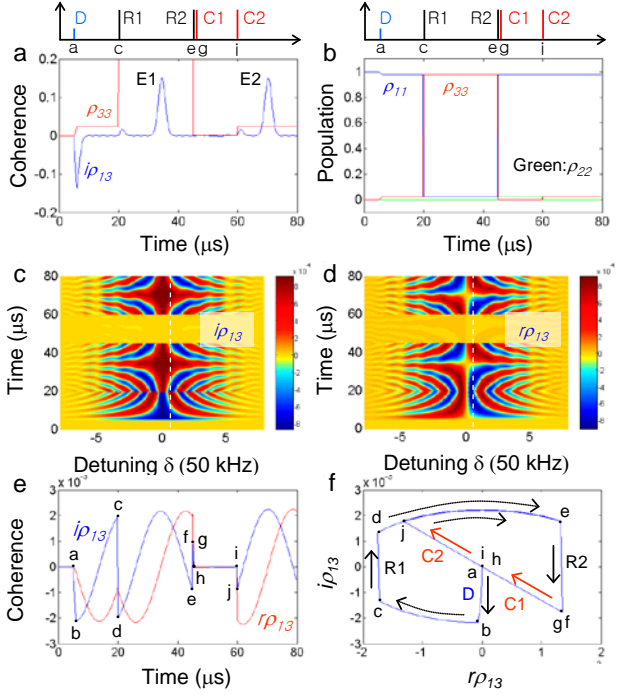

Figure 2. Numerical calculations for a single optical pulse storage. a, Double photon echoes with (E1) and without (E2) population inversion. Inset: Pulse sequence for Fig. 1. Each pulse area of R1, R2, C1, and C2 is $\pi$. The pulse area of D is $\pi/10$. The pulse duration of D is 1 µs. Pulse duration of R1, R2, C1 and C2 is 100 ns, respectively. b, Population in each state. c and d, Coherence as functions of time and atom detuning for (a). e, Coherence for an atom detuned by $\delta=30$ kHz: see the dashed lines in (c) and (d). Each letter indicates pulse timing of pulses in the inset of Fig. 1. f, Bloch vector notation for a detuned atom of Fig. 1 by $\delta=10$ kHz. Each pulse is a square pulse, and the interacting system of Fig. 1a is Gaussian distributed.

Figure 2 shows numerical calculations for the scheme of Fig. 1 for a single data pulse by solving nine time-dependent density matrix equations without any approximation, where optical decay constants are set at zero for simplification and visualization purposes. The insets at the top of Figs. 2a and 2b represent the corresponding optical pulse timing. Figure 2a shows echoes with population inversion (E1) and without inversion (E2), where the red curve denotes excited state population $\rho_{33}$. Here E1 is for the conventional two-pulse photon echoes, where the order of echo signals is reversed due to the rephasing by R1 (see Fig. 3), with forward propagation. The backward storage-time extended echo E2, however, is in the normal order without population inversion due to another rephasing by R2 along with control pulses C1 and C2. Because each rephasing pulse induces a backward evolution with a $\pi$ phase shift, the double rephasing by R1 and R2 results in a forward evolution with no phase change. However, the doubly rephased coherence is for absorption (see Fig. S2 of Supplementary information), explaining the need for the control deshelving pulses to deliver a $\pi$ phase shift for emission (see Figs. 2e and 2f). As shown in Fig. 2b, the R1 excited population inversion is swapped again by R2, where the resultant excited state population becomes the same as the initially excited state by the data pulse D. With C1 the excited state population is transferred from state $|3\rangle$ into state $|2\rangle$ (see the green line for $\rho_{22}$).

Figures 2c and 2d represent coherence evolutions in absorption and dispersion, respectively. As shown, the atom phase is reversed by R1 at t=20 µs and reversed again by R2 at t=45 µs resulting in echo absorption. However, with the contribution by C1 and C2, the final phase change turns out to be $\pi$ at t=60.2 µs, and normal echo type rephasing resumes. The interesting outcome is the forward coherence evolution direction, important for the second echo E2 to be nonreversed in the time domain (to be discussed in Fig. 3). Here the rephasing means a time reversed evolution across the rephasing pulse with a $\pi$ phase shift (see the color change in Fig. 3c, and color direction change in Fig. 3d at t=20 µs). The dashed lines are for a detuned atom group discussed for Figs. 2e and 2f.

In Fig. 2e, a detuned atom group (at 2 kHz bandwidth) is presented for coherence evolution. By data pulse D (a-b) optical coherence is excited by the absorption process, and then freely evolved at the speed determined by detuning $\delta$: $\exp(i\delta t)$. The first rephasing pulse R1 (c-d) swaps populations between states $|1\rangle$ and $|3\rangle$, and gives a $\pi$ phase shift to each $i\rho_{13}$ and $r\rho_{13}$: $\exp(i\delta t) \rightarrow \exp(-i\delta t)$. Here the evolution direction is backward, representing time reversal. This time reversal is the key mechanism of echo formation as a coherence burst. By the second rephasing pulse R2 (e-f) populations are re-swapped, and another $\pi$ phase shift is given to each atom: $\exp(-i\delta t) \rightarrow \exp(i\delta t)$. As a result, the evolution direction is forward resulting in echo absorption. The first control pulse C1 (g-h) transfers the excited state population $\rho_{33}$ into the spin state $|2\rangle$, where atom coherence becomes zero with another phase shift of $\pi/2$ (ref. 28). Then the second control pulse C2 (i-j) returns the transferred atoms back into state $|3\rangle$ with an additional phase shift of $\pi/2$ to resume the rephasing done by R2 but with a total $\pi$ phase shift in a forward evolution direction: $\exp(i\delta t) \rightarrow \exp(i\delta t)\exp(i\pi) = -\exp(i\delta t)$. Thus, the final state of the excited atoms at t=70 µs for echoes E2 is exactly the same as the initial state at t=6µs (position b) except for emission (see also Figs. 2c and 2d): $T_{E2}=T_{C2}+(T_{R2}-T_{E1})-\delta T$; $T_i$ stands for the arrival time of pulse $i$; $\delta T$ stands for the time delay of C1 from R1 (see Fig. S3 of Supplementary information). If $\delta T$ is larger than $(T_{R1}-T_{E1})$, no echo is generated because C1 functions rephasing halt[25,28]. Due to the double rephasing, the coherence evolution after the point j is exactly the same as that after the point b, both in the same order as the data pulses.

Figure 2f shows a Bloch vector evolution in a uv plane. As explained in Fig. 2e, R2 reinverts the atom's phase resulting in phasing with the data-excited coherence evolution, and the control pulses C1 and C2 give an additional $\pi$ phase shift for emission to echoes. Here an additional function of C1 via coherence transfer from state $|3>$ to $|2>$ is for storage-time extension dependent upon the spin dephasing time $\tau$ (no spin dephasing is assumed in the calculations). In $Pr^{3+}$ doped $Y_2SiO_5$ (Pr:YSO) the reported extended storage time is ~10 μs, which is determined by a spin inhomogeneous width of 30 kHz ($^3H_4 \pm 1/2 \leftrightarrow ^3H_4 \pm 3/2$) (refs. 5,25).

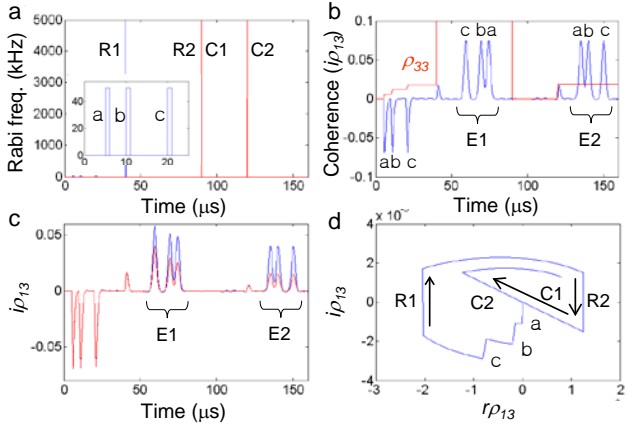

Figure 3. Numerical calculations for a multiple optical pulse storage. a, Coherence evolution. b, Phase versus time. $\Gamma_{21}=\gamma_{21}=0$. For the blue line, $\Gamma_{31}=\Gamma_{32}=1$ kHz; $\gamma_{31}=\gamma_{32}=2$ kHz. For the red line, $\Gamma_{31}=\Gamma_{32}=1$ kHz; $\gamma_{31}=\gamma_{32}=5$ kHz. Each pulse duration is 100 ns, except the data pulse at 1 μs. All other parameters are the same as in Fig. 2

Figure 3 shows numerical calculations for Fig. 1 with a multiple data pulse stream using experimental parameters of Pr:YSO. Figure 3a shows a pulse sequence with three consecutive a, b, and c data pulses. As shown in Fig. 3b, the photon echoes E1 are in a reversed order, while the photon echoes E2 are in the same order as the data pulses. More importantly, no atom population inversion occurs during E2 generation due to the double rephasing as discussed in Fig. 2 (see the red curve for $\rho_{33}$). Figure 3c shows echo decays under optical decoherence. As shown in both blue and red lines, the photon echo amplitude for E1 decays exponentially as a function of delay time of R1 from D, while the second photon echo E2 shows a flattened amplitude regardless of the delay. This flattened echo amplitude is a direct result of exact compensation between exponentially decreased coherence as a function of R2 delay and the exponentially increased coherence of the echo E1 for rephasing in a time-reversed manner. The magnitude of the flat echo amplitude is determined by the decay time to R2. However, in a slow decaying medium like Pr:YSO, the decoherence in Fig. 3c can be easily decreased by adjusting the optical pulse delay in a μs range, where the optical phase decay time is in the order of 100 μs, giving great advantage to quantum memory applications dealing with consecutive data photons of flying qubits. Figure 3d presents exactly the same pattern as in Fig. 2f, except for decoherence-induced amplitude reduction.

In conclusion a spontaneous emission-free photon echo quantum memory protocol was presented by all-optical control of atom phase. Unlike time reversed conventional two-pulse photon echoes, the present noise-free echo scheme keeps the same order as the data pulses due to double rephasing. Moreover a backward propagation scheme obtained by control deshelving pulses used for an additional $\pi$ phase shift to the double rephased atom ensemble, solves the inherent reabsorption dilemma in conventional two-pulse photon echoes. Finally, the noise-free flat echo amplitude does not suffer from exponential decay within the consecutive data photon stream. Photon storage time extension via a coherence conversion process by the control deshelving pulses is an additional bonus. Thus, the present scheme holds potential for quantum memories with nearly perfect retrieval efficiency, a multimode property storage-time extension, and no spontaneous or stimulated emission-caused quantum noises.